# When Integrated Photonics Should Be Wavy

Oliver R. Müller,[‡] J. J. Erik Maris,[‡] Yannik M. Glauser, Sander J. W. Vonk, and David J. Norris[*]

All authors: *Optical Materials Engineering Laboratory, Department of Mechanical and Process Engineering, ETH Zurich, 8092 Zurich, Switzerland*

[‡]These authors contributed equally to this work.

[*]Correspondence: dnorris@ethz.ch

## ABSTRACT

Increasing demand for high-performance optical devices drives the search for improved fabrication and design paradigms. While photonic circuits have traditionally used structures with two discrete "binary" height levels, grayscale "wavy" interfaces, or optical Fourier surfaces, have recently become possible. They provide precise control over the Fourier components that govern the optical response. This capability raises the question: When does a wavy device improve performance and why? Here, we show that wavy integrated processors exhibit superior accuracy and efficiency to binary analogs. Inverse-designed wavy interferometers have the design freedom to minimize outscattering to free space and backreflections. They reach a five-fold lower transmission error than binary counterparts and enable bandwidths up to 300 nm. We formalize these findings for other distinct integrated devices, such as photonic crystals, nanocavities, and beam emitters. A Fourier-optics analysis identifies a trade-off: wavy profiles manipulate light more accurately, whereas binary profiles excel in interaction strength. As such, tailored wavy profiles emit high-quality beams. However, binary profiles remain preferable in photonic crystals and nanocavities, where maximal index contrast is required and binarization-induced higher harmonics are benign. Thus, for quantum information, optical computing, and sensing, optical Fourier surfaces offer a route to miniaturized integrated circuits with improved performance.

## INTRODUCTION

Photonic integrated circuits (PICs) play a key role in the development of (quantum) optical technologies, including communication [1–3], optical computing [4–8], and sensing [3,9]. These circuits can be miniaturized using metastructures that shape light through an effective-permittivity landscape, defined by a height profile etched into a thin film of photonic material by using nanolithographic processes. Because traditional optical and electron-beam lithography can only remove this material to a fixed depth, these photonic structures typically contain "binary" height profiles (Figure 1a). Thus, the optimal design for a specific task must be adapted to be fabricable, which often compromises performance. For example, for in-plane manipulation of guided modes, scattering losses may be increased to ensure fabricability, which is problematic for quantum technologies where every photon counts [1–3,5–7]. For out-of-plane manipulation, spatial and angular control over the beam can be reduced, which is detrimental for trapped-ion quantum computing [4] and light detection and ranging (LiDAR) [9]. Overcoming the binary constraint could therefore reduce these losses and "sharpen" beam control, improving PIC functionality and lowering the energy consumed over the operational lifetime.

By relaxing the traditional, binary fabrication constraints, grayscale lithography promises a route to PIC components with improved performance. In particular, grayscale electron-beam lithography [10,11] and thermal scanning-probe lithography [12,13] have enabled the precise fabrication of photonic structures with "wavy" height profiles (Figure 1a), sometimes referred to as optical Fourier surfaces [14]. These techniques enable continuous control over the effective permittivity that the mode in the photonic layer experiences, conceptually similar to graded-index structures [15] and subwavelength gratings [16]. Because continuous control of the effective permittivity leads to more accurate light manipulation, an important question is: When does a wavy profile improve the performance of photonic integrated components? In this work, we investigate this question using Fourier optics combined with electromagnetic simulations and inverse design.

## RESULTS

### Diffraction regimes for optical Fourier surfaces

To explain the fundamental difference between wavy and binary integrated photonics, we begin by considering light propagating in the photonic layer. Such a guided mode can only couple efficiently to another mode (guided or free space) if their in-plane wavevectors $k_{||}$ are identical, i.e., they are phase-matched [17,18]. A mismatch in $k_{||}$ can be compensated by a periodic variation in the effective permittivity. This can be implemented by a periodic height profile (i.e., a grating) in the photonic layer. The grating contributes $\pm ng$ to $k_{||}$, where $n$ is the diffraction order and $g = 2\pi/\Lambda$ with grating pitch $\Lambda$ [19]. While sinusoidal gratings contain only a fundamental spatial frequency $g_1$, the corresponding binary (square-wave) gratings also contain higher harmonics, $g_3$, $g_5$, etc., which are odd multiples of $g_1$. These higher harmonics in binary structures complicate their ability to manipulate light, by opening additional coupling channels.

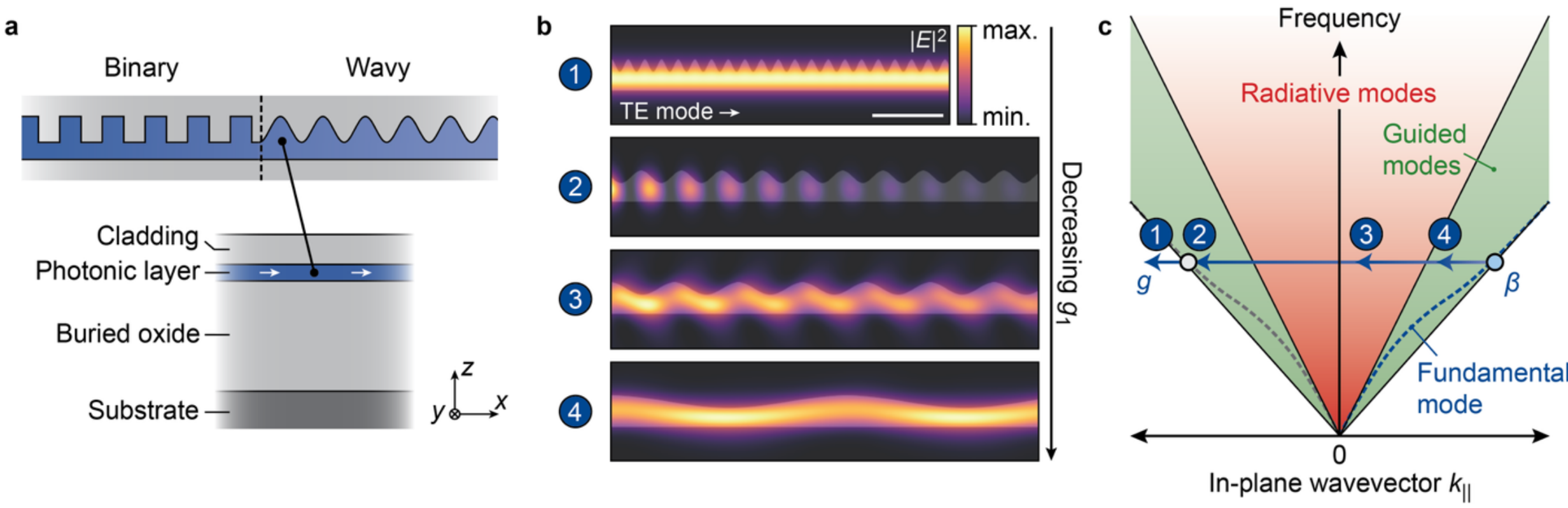

**Figure 1.** A Fourier-optics perspective on integrated photonics. (a) Side view of an integrated photonic chip containing height modulations. The layer stack consists of a photonic layer (e.g., Si) between a cladding and a buried oxide layer (both $SiO_2$) on a substrate (Si). Height profiles are etched into the photonic layer and can either be binary (two-level) or wavy (continuous). (b) Intensity distribution, $|E|^2$, of a transverse-electric (TE) fundamental mode at 1550 nm in a Si photonic layer with a sinusoidal grating of increasing pitch (decreasing $g_1$). At large and small $g_1$, depicted in (1) and (4), respectively, the light propagates with minimal coupling to other modes. In the intermediate regimes, shown in (2) and (3), the light is either reflected or outscattered to free space, respectively, thereby attenuating the forward-propagating mode. The scale bar is 0.5 µm. (c) Dispersion relation of the light confined in the photonic layer. A set of guided modes can only exist in the green region, which is bounded by the light lines in the cladding and photonic material. A continuous range of free-space modes exists in the light cone (red region), and coupling to these modes is observed as out-of-plane scattering. The dispersion curves of the fundamental waveguide modes are indicated by the dashed lines, where the filled blue dot marks a guided mode at a specific frequency. Coupling regimes, as in (b), are indicated by the arrows and numbers.

Consequently, sinusoids represent a natural basis to choose for diffraction phenomena. Figure 1b shows the simulated intensity of a fundamental mode with $k_{||} = \beta$ that is interacting with sinusoidal gratings of different periodicities (see Methods). We differentiate between four regimes with decreasing $g_1$, as depicted in the dispersion diagram in Figure 1c: (1) grating features are deeply subwavelength such that the structure behaves as a homogeneous effective medium, allowing lossless propagation; (2) guided forward- and counterpropagating modes are coupled by the grating, resulting in Bragg reflection; (3) guided modes couple to free-space radiation inside the light cone where $|k_{||}|$ is smaller than $2\pi n_{\mathrm{clad}}/\lambda_0$, leading to scattering losses (with $n_{\mathrm{clad}}$ the refractive index of the cladding material and $\lambda_0$ the free-space wavelength of the light); and (4) guided modes are only adiabatically perturbed by the large grating features, minimizing undesired coupling. To reduce optical losses in PIC components, spatial frequencies that enable coupling to counterpropagating modes (regime 2) and free-space modes (regime 3) must be avoided.

**Avoiding scattering losses with optical Fourier surfaces**

Using the regimes above, we now address the first advantage of wavy structures: reduced outscattering during propagation. Figure 2a plots the scattered power from finite-sized wavy and binary gratings as a function of their pitch $\Lambda$ (see Methods). When the grating is deeply subwavelength or fulfills the Bragg condition (regimes 1 and 2), the scattering losses are negligible. In contrast, when the mode couples into the light cone via first-order diffraction (regime 3), the scattering losses dominate. The important regime for reducing outscattering is regime 4, i.e., the pitch is much larger than the modal wavelength. Wavy gratings exhibit low scattering losses, especially compared to binary gratings of the same period (Figure 2a). This can be visualized by plotting the intensity of the propagating mode in the two structures (Figure 2b). Whereas, in the binary profile, standing waves and attenuation due to reflections and scattering, respectively, are clearly observed, the gradual height changes in the wavy structure allow better mode propagation.

Figure 2c,d further compare the binary and wavy gratings from Figure 2b. For each case, we plot the Fourier transforms of the height profile and the electric field. The latter is calculated from a line trace below the bottom of the photonic layer (horizontal dashed lines in the lower insets of Figure 2c,d) [20]. The plots reveal the Fourier components that are present in terms of $g$ for the height profiles and $k_{||}$ for the electric fields. A comparison of the top and bottom panels in Figure 2c,d shows that each $g$ component in the structure gives rise to a corresponding $k_{\parallel}$ component in the guided mode due to first-order coupling ($\beta \pm g_1$). In addition, higher-order interactions appear as additional $k_{\parallel}$ components ($\beta \pm 2g_1$, $\beta \pm 3g_1$, etc., for further discussion see Supporting Information, Sections S1.3 and S1.4).

For the binary grating (Figure 2c), the higher harmonics that are in the leaky region ($-g_3$, $-g_5$, and $-g_7$; red-shaded region) cause first-order coupling to free-space modes. Moreover, additional higher-order coupling terms ($\beta + g_1 - g_5, \beta - g_3 - g_1$, etc.) lead to further $k_{\parallel}$ components in the light cone, all spaced by $|g_1|$. This introduces undesired scattering losses from the binary grating. In contrast, the wavy grating (Figure 2d) does not contain the higher harmonics. Only weak higher-order diffraction related to $g_1$ ($\beta - 3g_1$ and $\beta - 4g_1$) causes scattering into the light cone.

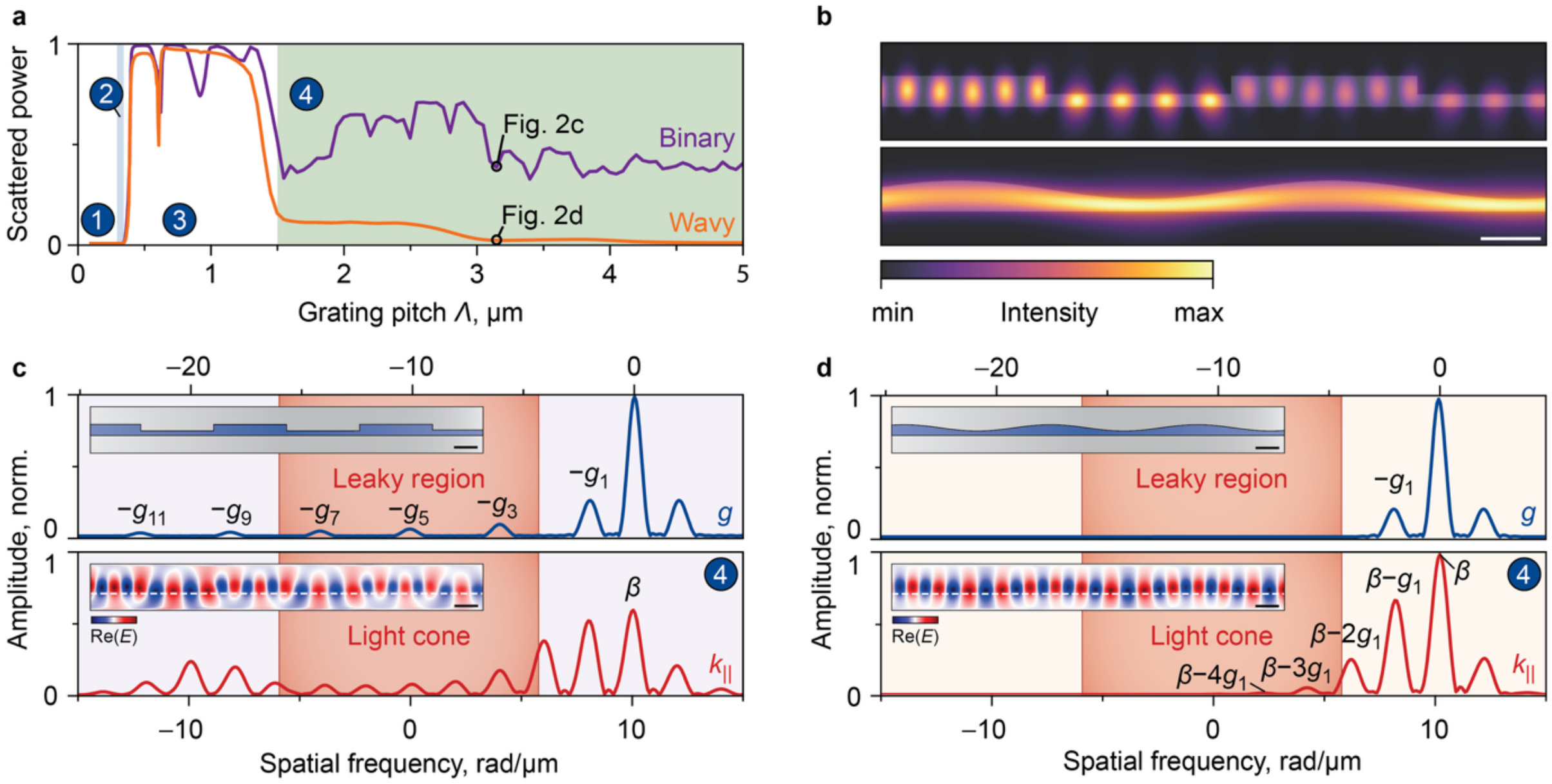

**Figure 2.** Coupling to free-space modes in wavy and binary gratings. (a) Simulated power of light scattered from binary (purple line) and wavy (orange line) gratings as a function of grating pitch for a corrugation depth of 60% of the photonic-layer thickness. The scattering regimes 1–4 (see Figure 1b,c) are indicated. In regimes 1 and 2, binary and wavy gratings do not outscatter light. In regime 3, both gratings outscatter strongly by first-order diffraction. In regime 4 (green region), outscattering from the binary gratings remains strong, whereas the wavy gratings show only weak losses. (b) Intensity distribution, $|E|^2$, of the fundamental TE mode in a grating with 3.15-µm pitch and 60% depth. The mode intensity in the binary grating is attenuated due to reflections and outscattering, whereas it remains high in the wavy grating. (c,d) Spatial frequencies of the height profile, $g$ (top panels), and of the electric fields, $k_{\parallel}$ (bottom panels), for the binary

and wavy gratings in (b). The zero in the spatial frequency of the structure has been aligned with the fundamental spatial frequency of the propagating mode, $\beta$. Peaks in the light cone (red region) can couple to free space. All scale bars are 0.5 µm.

This explains the differences between the purple and orange curves in Figure 2a for gratings with pitches larger than 1.5 µm (green-shaded region). Binary gratings always contain higher harmonics that lead to strong outscattering by first-order diffraction. For wavy gratings, weaker plateaus are observed that correspond to the higher diffraction orders. Scattering is sequentially reduced with increasing pitch as each higher diffraction order fails to reach the light cone (Figure 2d). Since the strength of the diffraction orders scales with grating depth, scattering can be further reduced for shallower gratings (see Supporting Information, Section S1).

These observations allow us to formulate two design rules for in-plane light manipulation: (1) wavy integrated photonic elements will have lower scattering losses than binary ones when they contain low spatial frequencies with $g_1$ too small to facilitate first-order coupling to the light cone; and (2) the amplitude of each of the spatial frequencies in the wavy height profile should be adjusted according to its potential to scatter into the light cone.

**Optical Fourier surfaces for inverse-designed PIC components**

Our design rules are particularly relevant for photonic inverse design, which provides a route to compact integrated devices with complex functionality [21,22]. In these methods, a figure of merit is minimized through optimization of the height profile for a photonic layer. A common approach is to perform gradient-descent optimization, using the adjoint method to obtain the gradients (see Methods and Supporting Information, Section S2). The process often provides designs with complicated and unintuitive height profiles [21–24], which contain a range of spatial frequencies.

To illustrate the benefit of wavy structures in inverse design, we demonstrate an integrated component with well-defined complex transmission coefficients between two input and two output waveguides. Such devices are generic linear photonic processors that can act as compact interferometers, beam splitters, matrix–vector multipliers, etc., depending on the chosen transmission matrix. Here, we target coefficients that lead to constructive and destructive interference in the top and bottom output waveguides, respectively. The first step in the design process of such an interferometer is to optimize the height profile without topography constraints. Because the gradients in the adjoint method are based on electromagnetic waves, this leads in general to a wavy topography (Figure 3a, top; see also further discussion below). Due to fabrication constraints, this design has then traditionally been binarized in a second step (Figure 3a, bottom; Supporting Information, Section S2.1). The desired interference in the outputs can be seen in the magnetic fields ($H_z$, Figure 3b) and the Fourier transforms of these fields reveal the $k_{||}$ components (Figure 3c). In the wavy device, the weak $k_{||}$ components inside the light cone indicate that the inverse-design process minimized scattering losses by suppressing spatial frequencies in the leaky region (Figure 3c, top). However, binarization inherently introduces higher harmonics of all $g_1$ components, leading to stronger outscattering (Figure 3c, bottom).

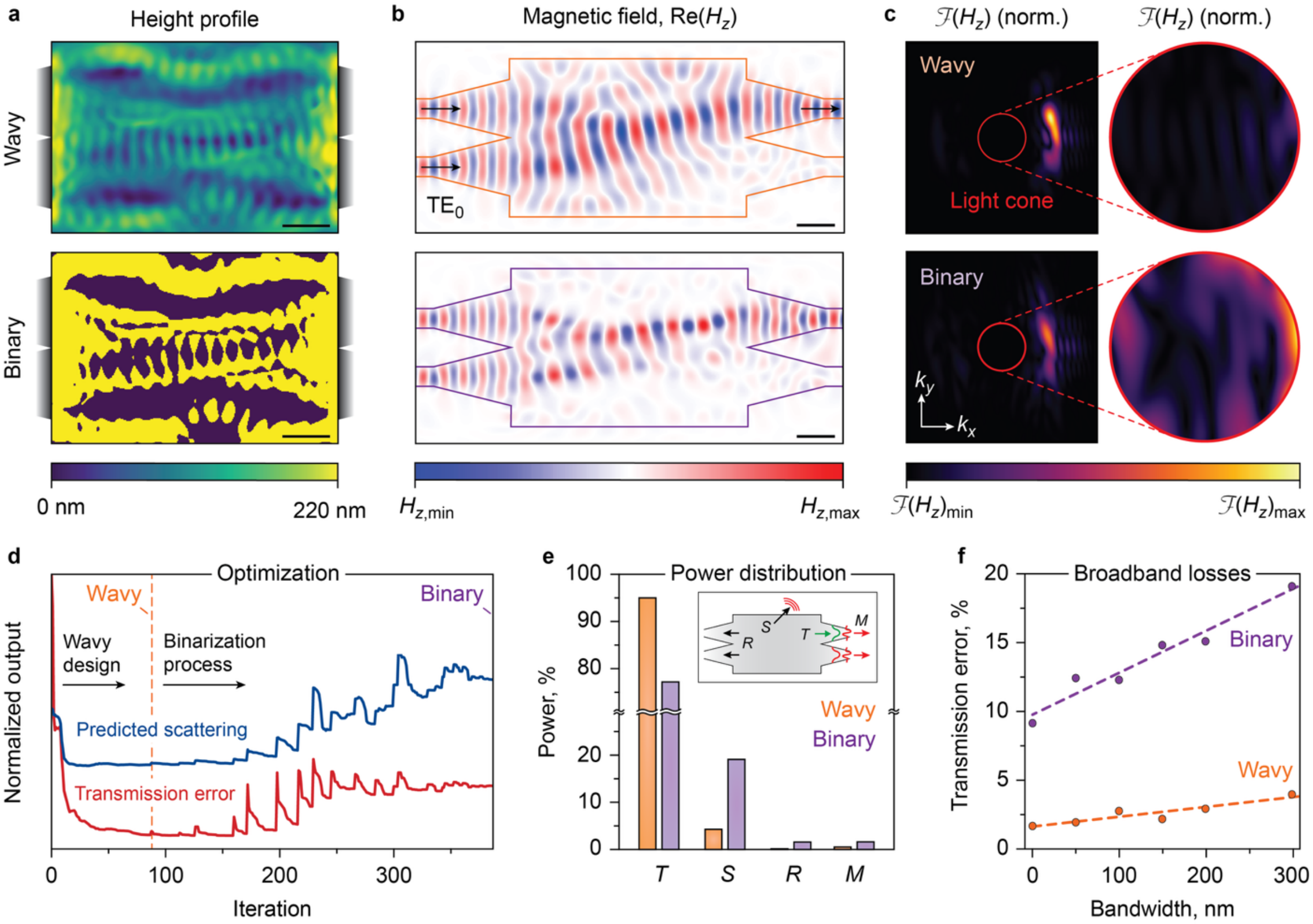

**Figure 3.** Inverse design of integrated photonic devices. (a) Height profile of a photonic interferometer before (wavy, top) and after (binary, bottom) the binarization process, optimized for a 100-nm bandwidth. (b) Magnetic field ($H_z$) of the TE modes injected in the input waveguides of the devices in (a). The interferometer should direct all power to the top waveguide on the right. Both fields are normalized to the same false color scale, and differences in field strengths are a result of interference. The scale bars are 1 µm.

(c) Two-dimensional Fourier transform of the field in (b). The wavy interferometer minimizes Fourier components in the light cone. (d) Comparison of the transmission error and predicted scattering [25] ($k_{||}$ integrated over the light cone) as a function of the iteration in the inverse-design optimization. (e) Optical-power distribution ($T$, transmission; $S$, outscattering; $R$, backreflection; $M$, mode conversion) for the optimized wavy (orange) and binary (purple) interferometers. (f) Transmission error of the wavy (orange) and binary (purple) interferometers optimized for a set of bandwidths centered at 1550 nm.

Figure 3d shows the evolution of the simulated transmission error (red curve) and the predicted outscattering (blue curve) as the optimization process proceeds. The latter is determined directly from the $k_{||}$ components inside the light cone [25]. For the wavy design (Figure 3d, left of dashed orange line), the transmission error is first minimized. Then, during the gradual binarization process, it spikes at the beginning of each epoch in the transition to a binary profile and fails to return to the prior optimum. Consequently, the final binary design has an increased transmission error. The correlation between the red and blue curves indicates that the loss in performance is caused by outscattering. This is consistent with Figure 3e, which reveals that scattering, $S$, dominates over other loss channels (backreflection, $R$, and mode conversion, $M$) in the binary interferometer.

In addition to lower transmission errors due to decreased outscattering, optimized wavy devices enable excellent broadband performance compared to binary devices (Figure 3f). We compared interferometers with spectral bandwidths up to 300 nm around the central wavelength of 1550 nm. The optimized wavy design exhibited transmission errors that were approximately 5× lower over the entire range. This improved broadband performance enables parallel processing via wavelength multiplexing, which is particularly interesting for optical-computing applications [26,27].

As mentioned above, the inverse-design process leads in general to wavy structures. However, the "degree of waviness" is affected by material and geometric parameters, which constrain the ability of the device to control the path of light. Two parameters are important: the effective-permittivity contrast and the device footprint [28,29]. The first, which is connected to the height variation and the bulk permittivity of the photonic layer, determines the interaction strength, while the second determines the interaction length. Given a sufficiently large effective-permittivity contrast or footprint, inverse design leads to wavy height profiles. If these parameters are restricted (i.e., low contrast and/or small footprint), the device may not provide the required interaction. The inverse-design process then pushes to the limits of the allowed effective-permittivity range (e.g., for the wavelength splitter in Figure 4a). Because this leads to steep edges in the profile, the coupling strength (amplitude of $g_1$) is in competition with outscattering due to the introduction of undesired spatial frequencies ($g_3$, $g_5$, etc.). The profile becomes wavy when the allowed permittivity contrast is increased (Figure 4b versus 4a) or the footprint of the device is enlarged (Figure 4c versus 4a). Thus, to achieve the best performance, the design must be given the freedom to be wavy.

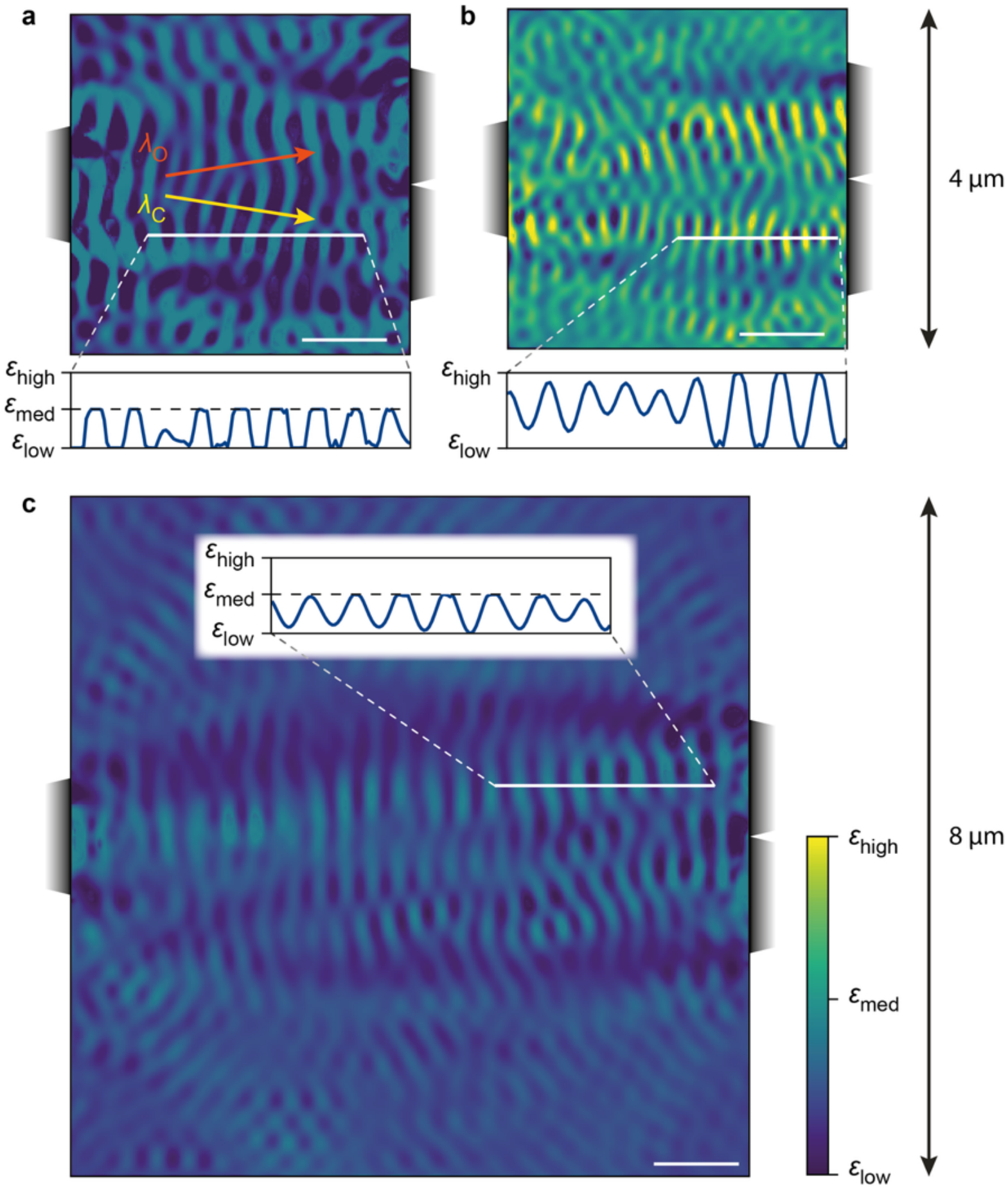

**Figure 4.** The "degree of waviness" of inverse-designed integrated components. (a) Effective-permittivity profile of an inverse-designed wavelength splitter (telecom O- and C-bands) under the constraints of a low effective-permittivity contrast and small device footprint. A crosscut (white line) of the effective permittivity is shown (bottom). Limiting both the contrast and the footprint leads to a "less wavy" design due to the reduced interaction strength and length. (b,c) Profiles of the wavelength splitters in (a) designed with the same footprint but larger contrast (b) and the same contrast but larger footprint (c). In both (b) and (c), the increased interaction strength or length, respectively, leads to "more wavy" designs. In (a) and (c),

the effective permittivities represent the layer stack in Figure 1a ranging from $\varepsilon_{\text{low}} = 1.44^2$ (fully etched, $SiO_2$ only) to $\varepsilon_{\text{med}} = 2.87^2$ (fundamental TE mode of the 220-nm Si photonic layer at 1550 nm). In (b), the range is extended to $\varepsilon_{\text{high}} = 4.0^2$, an arbitrary permittivity above bulk Si to illustrate the contrast dependence. The design areas are 4×4 μm² (a,b) and 8×8 μm² (c), and the scale bars are 1 μm.

## Optical Fourier surfaces for beam shaping

While above the goal was to optimize in-plane manipulation while avoiding unwanted outcoupling, we now address the second advantage of wavy structures: enhanced control over out-of-plane diffraction. By tailoring the $g_1$ components in the height profile that couple to desired $k_{||}$ components in the light cone, the diffracted field can be shaped precisely (Figure 5a). Under the weak-scattering approximation, the desired height profile is simply a superposition of sinusoidal gratings with the required spatial frequencies, and corresponding amplitudes and phases [13,14,17,30,31], i.e., a wavy surface. More specifically, these spatial frequencies should provide phase matching between the guided mode and the desired emitted beam. To obtain the required design, we backpropagate the targeted output field to the grating surface and account for the guided reference mode to determine the required height profile [30] (see Methods and Supporting Information, Section S3).

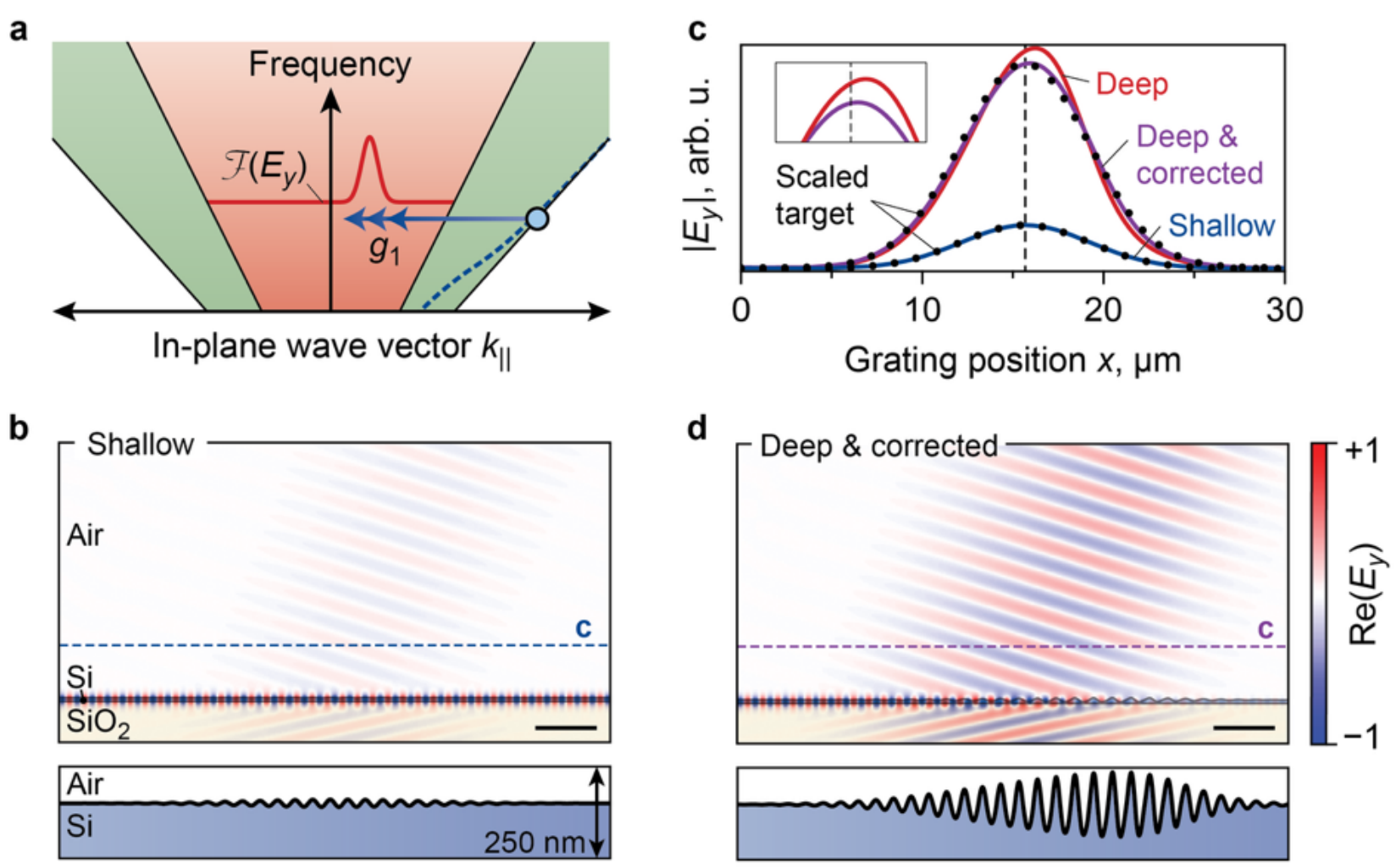

**Figure 5.** Beam shaping with wavy gratings. (a) Dispersion relation for outcoupling of light from a guided mode in a Si layer (dashed curve) to a targeted Gaussian beam (10-μm beam waist, 20° emission angle). The horizontal arrows indicate the $g_1$ components in the Fourier surface that allow the guided mode (blue dot) to couple to the Gaussian angular spectrum of the targeted beam (red curve). (b) The diffracted electric

field of a shallow grating with the required $g_1$ components. A close-up of the grating is given in the lower panel. (c) Amplitude of the electric field above the grating at the position of the dashed lines in (b) and (d) for shallow and deep gratings, respectively. To visualize beam-shape deviations, the targeted profile is scaled to the respective field amplitudes (black circles). When the deep gratings are corrected for local deviations, both the beam profile and field amplitude are optimized. (d) The diffracted electric field as in (b) for a deep grating corrected for local deviations. The scale bars are 5 μm.

We demonstrate the potential of Fourier surfaces for out-of-plane manipulation by designing outcouplers that create Gaussian beams at a 20° diffraction angle. Figure 5b shows an example device and its resulting emitted optical field. Because its height profile is shallow, the weak-scattering approximation applies, and the shape of the outcoupled beam can easily match the targeted Gaussian form. However, such shallow profiles also exhibit weak field amplitudes (blue curve, Figure 5c). Deeper profiles provide the necessary permittivity contrast to increase the output efficiency, but the diffracted beam shape then deviates from the target (red curve, Figure 5c). For such deep gratings, the height-dependent effective refractive index influences the phase of the guided mode, and the continuous outscattering depletes its amplitude. This causes a mismatch between the complex field in the waveguide and the reference wave assumed in the analytical design. By correcting for these deviations (see Methods and Supporting Information, Section S3), the targeted shape of the diffracted beam can be matched while maintaining high field amplitudes (purple curve, Figure 5c). Figure 5d plots the expected output from such an optimized device. These results demonstrate that a simple analytical design process that includes the corrections for local deviations can lead to wavy height profiles that provide efficient and high-quality outcouplers.

## DISCUSSION

To answer the question of when integrated photonic components should be wavy, we summarize two important effects of binarizing a height profile. First, it introduces higher harmonics of the fundamental Fourier components (with additional broadening in the case of finite structures). Second, binarization increases the amplitude of the fundamental Fourier components. While the former deteriorates the control over light due to unwanted coupling, the latter increases the coupling strength due to an enhanced permittivity contrast. Thus, a trade-off exists between the accuracy and the strength of the light manipulation. This trade-off governs whether the height profile should be wavy or binary.

In Figure 6, we illustrate this accuracy–strength trade-off with four relevant nanophotonic structures: (i) photonic crystals, (ii) photonic crystal nanocavities, (iii) beam-shaping outcouplers, and (iv) integrated processors. (i) Photonic crystals (Figure 6a) favor binary profiles because their optical response benefits from maximal index contrast. Their higher harmonics remain subwavelength and thus do not affect their functionality. (ii) Photonic crystal nanocavities (Figure 6b) similarly benefit from the contrast of binary structures. Although the finite size of the cavity can introduce Fourier components in the leaky region that couple to radiative modes, apodization of the cavity can smooth the mode profile, minimizing outscattering [20,32]. This strategy provides sufficient mode control while maintaining strong interactions.

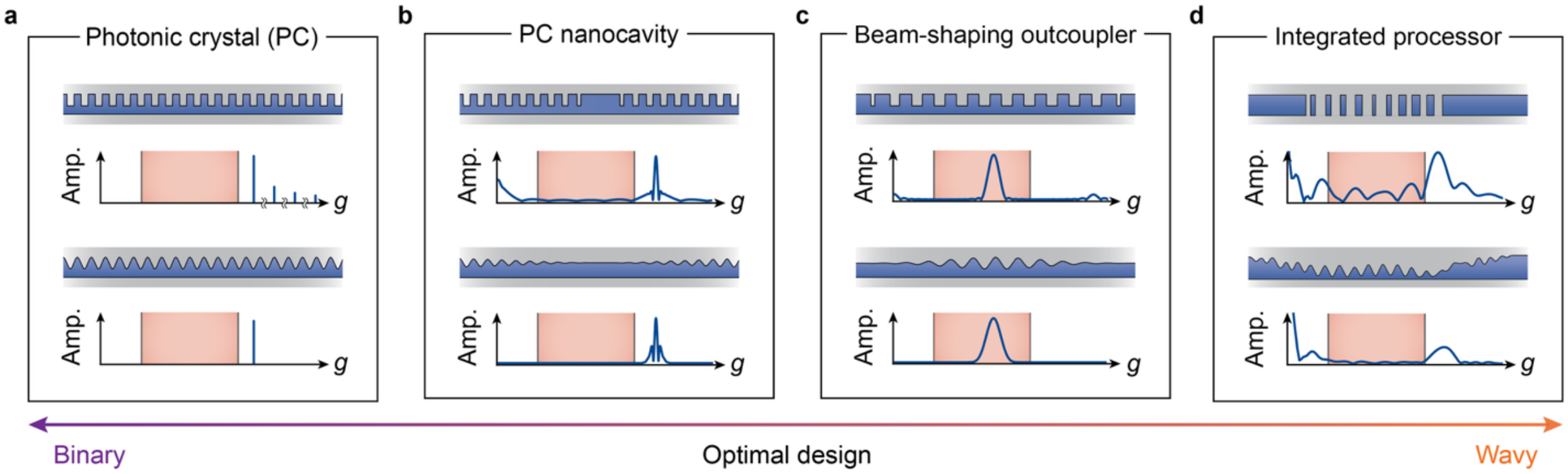

**Figure 6.** When should integrated photonics be wavy? Binary versus wavy versions of four nanophotonic structures along with their Fourier spectra. (a) Photonic crystals should be binary because they benefit from larger fundamental Fourier component(s) compared to wavy designs, while higher harmonics remain deeply subwavelength and therefore evanescent. (b) Similarly, photonic crystal nanocavities should also be binary, but leaky Fourier components must be suppressed by apodization. (c) Beam-shaping outcouplers should be wavy, as this provides precise control over Fourier components in the leaky region and enables efficient outcoupling of high-quality beams. (d) Inverse-designed photonic processors should similarly be wavy to reduce Fourier components in the leaky region and obtain precise, broadband in-plane light manipulation. Panels (a) and (b) show sketched structures with their calculated Fourier transforms. Panels (c) and (d) correspond to the devices designed and simulated in this work: (c) shows the structures schematically, and (d) shows crosscuts of the interferometers in Figure 3.

In both of these well-established structures, the coupling strength drives the trade-off, and our conclusions follow directly from the Fourier picture. In the remaining two, which we design and simulate explicitly above, accuracy becomes more important. (iii) Beam-shaping outcouplers (Figure 6c) target smooth, accurate field profiles. While a broadening in the Fourier spectrum of binary couplers complicates their optical response, its absence in wavy couplers enables precise tailoring of the leaky components. We demonstrated this in Figure 5 with a Gaussian beam emitter, designed via a straightforward Fourier-optics-based method including corrections for height-dependent phase accumulation and mode depletion. The emitted beam exhibits a smooth field profile, excellent agreement with the target Gaussian, and high coupling efficiency. (iv) Inverse-designed integrated processors (Figure 6d) rely most heavily on precise Fourier control. These structures shape the amplitude and phase of a guided mode with a broad range of spatial frequencies, including ones below the leaky region. When such height profiles are binarized, the low spatial frequencies generate higher harmonics that fall inside the leaky region. Our inverse-designed interferometers demonstrate this effect: wavy designs achieve higher efficiencies, while binarization increases outscattering. Moreover, the smooth profiles also minimize reflections and resonances, yielding excellent broadband performance.

## CONCLUSION

Wavy height profiles trade reduced refractive-index contrast for continuous, precise control over the effective-permittivity landscape. With the freedom to tailor the Fourier components of a height profile independently, optical Fourier surfaces enable low-loss broadband devices on a minimal footprint as well as accurate beam emitters. Thus, optical Fourier surfaces unify in-plane and out-of-plane light manipulation in a single physical framework. Where grayscale lithography is unavailable, binary profiles with subwavelength features approximate continuous control over the effective permittivity (albeit with some practical implications) [16]. Binary profiles remain preferable where maximal index contrast is required and the accuracy penalty is small or absent. More broadly, our study reframes binary fabrication constraints as a design choice rather than a fixed limitation. Advances in grayscale lithography and PIC design methods, combined with the design rules established here, offer a clear route toward the next generation of miniaturized integrated circuits for quantum information, optical computing, and sensing applications.

## METHODS

### Grating model

The diffraction losses of the wavy and the binary gratings were simulated using a finite-difference time-domain solver (FDTD, Ansys Lumerical 2022R2). The geometry of the grating assumed a waveguide with a 220-nm silicon photonic layer ($n = 3.48$) and a homogeneous silicon-dioxide cladding ($n = 1.44$). The grating pitch was varied from 100 to 5000 nm, and the grating depth from 10 to 100% of the photonic-layer thickness. The total grating length was 18 µm with apodizations of 5 µm on both sides. A two-dimensional crosscut of the grating in the ($x$, $z$)-plane was used to simulate a fundamental transverse-electric mode with a free-space wavelength of $\lambda_0 = 1550$ nm propagating through the structure. The outscattered radiation was recorded above and below the grating using frequency-domain monitors, and the reflected and transmitted light on the left and right of the grating, respectively, using mode-expansion monitors. The near field was captured 80 nm below the grating for the plane-wave decomposition of the modal field (dashed lines in Figure 2c,d). For further details of the grating model see Supporting Information, Section S1.

### Inverse design with the adjoint method

The interferometer and wavelength splitter were inverse designed with density-based topology optimization using the adjoint method [22–24]. To this end, a figure of merit was minimized via a gradient descent, performed by the L-BFGS-B solver from SciPy. The gradients were computed with the adjoint method using sets of electromagnetic simulations of the photonic device (FDTD, Ansys Lumerical 2022R2). The wavy devices were designed in two consecutive steps: (i) an unconstrained optimization, and (ii) a postprocessing optimization where a top-hat convolution filter with a radius of 70 nm was continuously applied to remove small spatial features. The binary devices were designed by taking the wavy profiles as initial guesses and applying a gradual binarization procedure, in which the design parameters were projected through a hyperbolic-tangent function of increasing steepness over successive epochs. This procedure converged to nearly binary designs, which were fully binarized by application of a threshold.

The interferometers were optimized in 3D to perform 2×2 matrix–vector multiplications on the complex amplitude of the two input modes. The target transmission matrix was

$$\boldsymbol{t}^{\mathrm{tar}} = \frac{1}{\sqrt{2}}\begin{pmatrix} 1 & 1 \\ \mathrm{i} & -\mathrm{i} \end{pmatrix},$$

which interferes the two (in-phase) input modes constructively in the upper output and destructively in the lower one. The figure of merit was defined as [33]

$$f_{\mathrm{ifm}} = \frac{1}{4N}\left(\frac{1}{\lambda_{\mathrm{max}}-\lambda_{\mathrm{min}}}\int_{\lambda_{\mathrm{min}}}^{\lambda_{\mathrm{max}}} \|w_0(\lambda)\,\boldsymbol{t}(\lambda) - \boldsymbol{t}^{\mathrm{tar}}(\lambda)\|_F^4 \,\mathrm{d}\lambda\right)^{1/2},$$

where $\|\ldots\|_F$ is the Frobenius norm, $\boldsymbol{t}$ the simulated complex transmission matrix, $N$ the size of $\boldsymbol{t}$, and $w_0$ a variable phase factor that applies a single phase shift to all elements in $\boldsymbol{t}$ such that $f_{\mathrm{ifm}}$ was minimized. The figure of merit was optimized over a wavelength range defined by $\lambda_{\mathrm{min}} = \lambda_{\mathrm{c}} - \lambda_{\mathrm{B}}/2$ and $\lambda_{\mathrm{max}} = \lambda_{\mathrm{c}} + \lambda_{\mathrm{B}}/2$ with $\lambda_{\mathrm{c}} = 1550$ nm and $\lambda_{\mathrm{B}}$ the bandwidth.

The wavelength splitters were optimized in 2D to split input modes with wavelengths in the telecom O- and C-bands into the upper and the lower output, respectively. The figure of merit was defined as

$$f_{\mathrm{WS}} = \frac{1}{2}\left|T_1(\lambda_{\mathrm{O}}) - T_1^{\mathrm{tar}}(\lambda_{\mathrm{O}})\right| + \frac{1}{2}\left|T_2(\lambda_{\mathrm{C}}) - T_2^{\mathrm{tar}}(\lambda_{\mathrm{C}})\right|,$$

where $T_{\mathrm{m}}$ is the transmission of the device and $T_{\mathrm{m}}^{\mathrm{tar}}$ the target transmission into the $m$th output, $\lambda_{\mathrm{O}} = 1305$ nm and $\lambda_{\mathrm{c}} = 1550$ nm are the central wavelengths of the telecom O- and C-bands, respectively. For further details about inverse design see Supporting Information, Section S2.

**Analytical design of wavy beam emitters**

The wavy beam emitters were simulated using a free-space wavelength of 1550 nm and a silicon-on-insulator layer stack with a 150-nm Si ($n = 3.48$) photonic layer (unperturbed waveguide), modulated up to 250 nm, on an $SiO_2$ ($n = 1.44$) buried oxide and an air cladding. The height profiles were designed using the analytical Fourier-optics-based approach that we recently reported [30] and two corrections that account for the phase and the amplitude mismatch in deeper gratings.

For a desired scattered field $E_{\mathrm{s}}(x)$ at the waveguide surface, the required real-valued height profile is

$$h(x) \propto \mathrm{Im}[E_{\mathrm{s}}(x)/E_{\mathrm{g}}(x)],$$

where $E_g(x)$ is the guided reference wave. Standard free-space backpropagation via the angular-spectrum method was used to obtain $E_s(x)$ for a target field specified at a desired output plane. Here, we targeted a Gaussian beam with a 10-µm beam waist emitted at 20° with respect to the surface normal.

For shallow gratings (Figure 5b), the reference wave is approximated by the unperturbed waveguide mode,

$$E_g(x) = u_g \exp(\mathrm{i}\, k_0 n_{wg}\, x),$$

where $n_{wg}$ is the effective refractive index of the unperturbed waveguide, and $u_g$ is the guided-mode amplitude, which is unity for an unperturbed waveguide. For deeper gratings (Figure 5d), we accounted for the height-dependent optical path length (OPL), and thus the phase accumulation, via $n_{eff}[h(x)]$ by re-mapping the longitudinal coordinates $x$ to $x_{OPL}$ according to

$$x_{OPL}(x) = \int_0^x \left\{\frac{n_{wg}}{n_{eff}[h(u)]}\right\} \mathrm{d}u,$$

where $n_{eff}[h(u)]$ was obtained by solving the self-consistency condition for the fundamental TE mode of a slab waveguide of thickness $h$. The height values were re-mapped to $x_{OPL}$ and subsequently resampled onto the original uniform spatial grid.

We additionally accounted for depletion of the guided mode due to outscattering. For a prescribed total outcoupling efficiency $\eta$, the guided-mode amplitude envelope was taken as

$$u_g(x) = \sqrt{1 - \eta \left[\int_0^x I_s(x')\, dx' \,/\, \int_0^L I_s(x')\, dx'\right]},$$

where $L$ is the total grating length, and $I_s = |E_s|^2$ is the scattered intensity at position $x$. This envelope was included in the reference wave used to calculate the final height profile.

## ADDITIONAL INFORMATION

### Supporting Information

The Supporting Information cited throughout this manuscript is not included in this preprint and will be made available alongside the journal publication.

## Author contributions

O.R.M. and J.J.E.M. contributed equally to this work. O.R.M. and J.J.E.M. conducted the optical designs, numerical simulations, data analysis, and interpretation of the results. Y.M.G. and S.J.W.V. contributed conceptually to the work and to the preparation of the manuscript. D.J.N. initiated and supervised the project. O.R.M., J.J.E.M., and D.J.N. wrote the manuscript with contributions from all authors. All authors discussed the results and contributed to the paper.

## Conflict of interest

The authors declare the following competing financial interest: Y.M.G., S.J.W.V., and D.J.N. have filed a patent application related to the linear design of optical Fourier surfaces presented in this study.

## Data availability

All the data and methods needed to evaluate the conclusions of this work are presented in the main text and the Supporting Information. Additional data and code can be requested from the corresponding author.

## Acknowledgments

This project was supported by the Swiss National Science Foundation (SNSF) under grant no. 2000-1-240090. S.J.W.V. acknowledges funding from SNSF under grant no. 200021-232257.